# Hierarchical porosity inherited by natural sources affects the mechanical and biological behaviour of bone scaffolds

Simone Sprio[1*], Silvia Panseri[1], Monica Montesi[1], Massimiliano Dapporto[1], Andrea Ruffini[1], Samuele M. Dozio[1], Riccardo Cavuoto[2], Diego Misseroni[2], Marco Paggi[3], Davide Bigoni[2], Anna Tampieri[1]

[1]Institute of Science and Technology for Ceramics, National Research Council, Via Granarolo 64, 48018 Faenza, Italy
[2]University of Trento, Department of Civil, Environmental and Mechanical Engineering, Via Mesiano 77, 33123 Trento, Italy
[3]IMT School for Advanced Studies Lucca, Piazza San Francesco 19, 55100 Lucca, Italy

*Corresponding Author:
Simone Sprio
e-mail: simone.sprio@istec.cnr.it
Tel. +39 0546 699759            Fax. +39 0546 46381

**ABSTRACT**
A 3-D porous apatite scaffold (B-HA), recently obtained through biomorphic transformation of a natural wood, is investigated on its multi-scale porous structure determining superior mechanical properties and biological behaviour. B-HA shows hierarchical pore architecture with wide aligned channels interconnected with smaller tubules, thus recapitulating in detail the lymphatic network of the original wood template. As induced by its biomimetic architecture, B-HA displays values of compression and tensile strength and stiffness, higher than the values usually measured in sintered ceramics with isotropic porosity. Furthermore, B-HA shows a ductility not common for a pure ceramic body and a tensile strength higher than its compression strength, thus occupying a zone in the Ashby chart where ceramics are usually not present. Cell co-culture tests in bioreactor report encouraging results in enhancing the complex tissue regeneration process, thus making B-HA very promising as a scaffold able to promote bone regeneration, particularly for large bone defects.

**Declaration of interest**: none

## 1. INTRODUCTION

In the development of bone scaffolds, it is widely accepted today that chemical composition, morphology of pores and mechanical response of the scaffold are all relevant aspects to induce appropriate cell response, which is pivotal to achieve new bone tissue formation and remodeling [1, 2]. The process of new bone formation has to be always accompanied by the growth of a vascular network, critical for the effective distribution of oxygen and nutrients, as well as for the removal of waste and metabolic sub-products [3]. To tackle this goal, extensive research has been dedicated to the study of compositional, morphological and mechanical aspects of bone scaffolds favoring the complex tissue regeneration process[4, 5]. However, insufficient vascularization still remains a critical problem, particularly when large bone defects have to be treated [6]. In fact, in the treatment of critical size defects new bone penetration is often limited to a few millimeters, because the lack of effective nutrient supply in the inner regions of the scaffold provokes bone necrosis or insufficient cell penetration [7, 8].

Pore size, morphology, interconnection and mechanics are all features directly influencing the capability of fluid flowing, cell penetration and development of new bone and vessels [9]. More in detail, the presence of channel-like porosity is considered as extremely important for vascular development, as it mimics the pore organization of the osteon structures that in long bones host the vascular network [10]. From the mechanical viewpoint, the scaffold architecture directly influences the strength, since the ones exhibiting multi-scale porosity can induce tortuous fracture growth, thus positively influencing toughness. Indeed, the development of bioactive materials matching the exceptional mechanical performance of bone and its vascular tree has been intensively pursued [10-14]. The closest approaches to a sinter-free process ensuring the maintenance of a bioactive composition and the presence of relevant ions in the apatite structure [15-17] are: i) the use of a freeze casting process [18], ii) the use of 3-D printing technologies making use of self-hardening ceramic-based pastes [19, 20].

In the case of option i) the use of bioerodible polymers is required; however, they are affected by degradation profiles often mediated by hydrolytic reactions, rather than by cell metabolic activity[21], thus jeopardizing the formation of mature bone tissue [22, 23]. When the option ii) is adopted, 3-D scaffolds with ineffective mechanical performance are obtained, due to the difficulty in controlling the microstructure development during the setting process [24, 25].

The present article aims to investigate the relationship existing among the multi-scale structure and organized channel-like morphology vs. the mechanical performance and biological behavior of a biomorphic ceramic scaffold, given its high bioactive chemical composition following a sintering-free biomorphic synthesis [26]. The role played by the multi-scale hierarchical microstructure on the mesenchymal and endothelial cell fate is also explored in a dynamic 3-D environment. Such experiment is performed in the presence of co-cultured human mesenchymal stem cells (hMSCs) and human umbilical vein endothelial cells (HUVECs). The mechanical and biological performances of the new scaffold are found superior than those of sintered hydroxyapatite-based blocks, characterized by a randomly organized macro-porous structure. Besides, the scaffold shows enhanced biological effects, as induced by the unique 3D biomimetic architecture.

## 2. MATERIALS AND METHODS

*2.1 Preparation of the scaffolds*

The apatite scaffold (hereinafter coded as B-HA) was prepared following a method described elsewhere [26]. Briefly, cylindrical wood pieces (*Calamus Manna*) were subjected to pyrolysis under nitrogen gas flow, then transformed into a biomorphic apatite body presenting multiple doping with $Mg^{2+}$ and $Sr^{2+}$ ions through a sequence of heterogeneous reactions with gaseous reactants at supercritical conditions, concluded by a hydrothermal process carried out at 220 °C. As a comparison material for the biological and mechanical tests, a commercial sintered calcium phosphate body with similar phase composition and porosity extent was used (Kasios®, Atoll, hereinafter coded as S-HA)[27].

*2.2 Physicochemical characterization*

Physicochemical characterization was carried out on samples crushed into fine powders by mortar and pestle. The phase composition was obtained by X-ray diffraction (XRD), using a D8 Advance Diffractometer (Bruker Karlsruhe, Germany) equipped with a Lynx-eye position sensitive device (CuKα radiation: λ = 1.54178 Å). XRD spectra were recorded in the 2θ range from 20 to 60° 2θ with a counting time of 0.5s and a step size of 0.02°. Chemical analysis was performed on dried samples using ICP-OES spectrometer (Agilent 5100). About 20 mg of the sampling material was dissolved into 2 ml of nitric acid (Sigma Aldrich 65 vol%) then diluted with milliQ water to obtain 100 ml of solution. The solution was then analysed using standard prepared from primary standards (1000 ppm, Fluka).

*2.3 Morphological characterization*

Field emission gun scanning electron microscopy (FEG-SEM) (Sigma NTS GmbH, Carl Zeiss, Oberkochen, Germany) was used to evaluate the morphology of the final material at the multi-scale. The samples were placed on an aluminum stub and covered with a thin layer of gold to improve conductivity. Optical microscopy in reflected light illumination (Olympus MIC-D Digital Microscope, incorporating cutting-edge light-emitting diode) was used to investigate the structure of native wood. The open and total porosity of the studied ceramics was measured by Archimedes' method and geometrical weight-volume evaluation, respectively. The specific surface area (SSA) of the scaffold was measured by nitrogen adsorption method, following the Brunauer–Emmett–Teller (BET) model (Sorpty 1750, Carlo Erba, Milan, Italy).

*2.4 Mechanical characterization*

In order to investigate the mechanical properties of B-HA, 58 specimens of different shape were tested under different loading conditions. For comparison, tests on S-HA samples were also carried out. Mechanical testing was conducted with a Midi 10 and a Beta 100 electromechanical testing machines (Messphysik Materials Testing) used in different configurations for uniaxial compression and three-point bending. The mechanical tests on B-HA were carried out along the main channel direction. Furthermore, crack propagation was investigated with additional compression tests conducted inside a scanning electron microscope (Zeiss EVO MA15 equipped by the tensile/compressive stage DEBEN 5000S). Further morphological investigation of crack patterns at higher resolution have been conducted on fractured specimens using the field emission gun scanning electron microscope (FEG-SEM) Tescan GAIA3.

Prismatic, cylindrical and tubular samples have been used to assess the compressive strength and stiffness of B-HA. The bases of the specimens were smoothed before testing, to reduce effects of uneven contacts at the specimen/machine surface. The samples were placed in direct contact with two circular steel plates (40mm in diameter and 5mm thick) and compressed, by imposing the displacement of the upper plate, until failure occurred.

Three-point bending tests have been performed on prismatic samples only, due to incompatibility of cylindrical and tubular shapes with the test set-up.

Ultrasonic testing was conducted to obtain the elastic stiffness of the material on a scale closer to that typical of bone and vessel cells. The tests were performed by measuring the flight time (corresponding to the time when the sensor reads a response of 80% of the applied amplitude) needed for compressive wave to travel along the whole specimen. P-wave transducers with characteristic frequencies of 0.5 MHz and 1 MHz were used[28].

*2.5 In vitro cell cultures*

Human adipose tissue-derived stem cells (hADSCs, ATCC) and human umbilical vascular endothelial cells (HUVECs, ATCC) were used for the biological study. In detail hADSCs were cultured in αMEM Glutamax (Gibco), containing 15% Fetal Bovine Serum (FBS, Gibco) and 1% penicillin-streptomycin (100 U/ml-100 µg/mL, Gibco), 10 ng/mL FGF. The cell culture was kept at 37 °C in an atmosphere of 5% $CO_2$. HUVECs were cultured in vascular cell basal medium (ATCC) supplemented with endothelial cell growth kit-VEGF (ATCC). Cells were detached from culture flasks by trypsinization, centrifuged and re-suspended. Cell number and viability were assessed with the trypan-blue dye exclusion test. Each sample (diameter 8 mm, height 4 mm), sterilized by 25 kGy γ-ray radiation prior to use, was placed one per well in a 24-well plate and pre-soaked in culture medium for 72 h at 37 °C. The U-CUP perfusion bioreactor system (Cellec Biotek AG) was used for the co-culture study. Briefly, hADSCs were seeded at $1.0 \times 10^6$ cells/scaffold with a bidirectional flow rate of 3 ml/min for 18 h, then all the media were collected, and the seeding efficiency was evaluated by counting the cells number left in the culture media after trypan-blue staining. The cell-seeded constructs were then cultured with a bidirectional perfusion flow rate of 0.3 ml/min in osteogenic medium (αMEM Glutamax containing 10% FBS and 1% penicillin-streptomycin (100 U/ml-100 µg/mL, Gibco), 10 mM β glycerophosphate, 50 µg/ml ascorbic acid, 100 nM dexamethasone)[29]. After 10 days, HUVECs were seeded ($1.0 \times 10^6$ cells/scaffold) following the same procedure mentioned above and the co-cultured constructs were left for 10 additional days. A ratio 1:1 of osteogenic culture medium and vascular cell basal medium supplemented with endothelial cell growth kit-VEGF was used for all the co-culture studies. The medium was changed twice a week. All the cell-handling procedures were performed in a sterile laminar flow hood. All cell-culture incubation steps were performed at 37 °C with 5% $CO_2$. S-HA bodies were used as control group.

*2.6 Histological analysis*

B-HA samples were placed in plastic vial, fixed in formalin 4%, dehydrated and degreased under vacuum with a series of alcohol washes: 1 hour of 70%, 80%, 95% and 100% ethanol, another 100% ethanol overnight. The infiltration was performed starting from two washes of 1 hour each with propylene oxide, followed by a 1:1 mixture ratio of propylene oxide and Epon E812 resin for 1 hour, followed by an overnight step with a 1:2 ratio of the same mixture. Finally, a 2-hour step with pure Epon E812 resin was performed and then replaced with new Epon E812 resin for the curing step performed at 60 °C for 24 hours. The cured resin block was freed by the plastic vial and cut producing serial 10 μm thick slides using a microtome (MRS 3500, Histo-Line Laboratories, Pantigliate, Milano, Italy) with a tungsten blade (IC1567, Delaware Diamond Knives Inc., Lancaster Pike, Wilmington, Delaware, USA). The sample sections were placed on gelatin-coated slides for further analysis. B-HA slides were processed for hematoxylin and eosin staining; slides were placed in a coupling jar and washed with Milli-Q $H_2O$. The slides were then dip in Mayer's hematoxylin for 7 minutes at RT. Subsequently the slides were rinsed in flowing water for 10 minutes and washed briefly in Milli-Q $H_2O$. The eosin Y solution was added and kept for 1 minute; the excess was then washed with Milli-Q $H_2O$. The slides were mounted with an aqueous mounting medium and covered with a coverslip. Images of each sample slices were acquired in bright field with a Digital Sight DS-Vi1 camera (Nikon) mounted on the microscope (Nikon Ti-E).

*2.7 Quantitative real-time polymerase chain reaction (q-PCR)*

At day 20, cells grown on the S-HA samples, used as control (CT), were homogenized and total RNA extraction was performed by use of the Tri Reagent, followed by the Direct-zol™ RNA MiniPrep kit (Euroclone) kit according to manufacturer's instructions. RNA integrity and quantification were analyses by NanoDrop™ One Microvolume UV-Vis Spectrophotometers (Thermo Scientific) following manufacturer's instructions. Total RNA (500 ng) was reverse transcribed to cDNA using the High-Capacity cDNA Reverse Transcription Kit, according to manufacturer's instructions. Quantification of gene expression, using Taqman assays (Applied Biosystems), for Runt-related transcription factor 2 (Runx2, HS00231692_m1), Alkaline phosphatase (ALP, HS01029144_m1), Integrin binding sialoprotein (IBSP, HS00173720_m1), Collagen 1 (COL1A1 Hs00164004_m1), Von Willebrand factor (VWF, Hs01109446_m1), Platelet And Endothelial Cell Adhesion Molecule 1 (PECAM1, Hs01065282_m1), and glyceraldehyde 3-phosphate dehydrogenase, used as housekeeping gene, (GAPDH, Hs99999905_m1) were performed by use of the StepOne™ Real-Time PCR System (Applied Biosystems). N. 2 scaffolds for each group were analysed, using three technical replicates for each experiment. Data were collected using the OneStep Software (v.2.2.2) and relative quantification was performed using the comparative threshold (Ct) method (ΔΔCt), where relative gene expression level equals $2^{-\Delta\Delta Ct}$ [30].

*2.8 Immunofluorescence analysis*

The cell seeded constructs were fixed in 4% (w/v) paraformaldehyde, blocked with 20% normal goat serum and permeabilized with 0.1% (v/v) Triton x-100. The samples were incubated overnight at 4 °C, with antibodies anti-Osteonectin, anti-IBSP, anti-VWF and anti-PECAM1 (Abcam) followed by incubation with secondary antibodies Alexa Fluor 488 goat anti-rabbit (Molecular Probes) and Cy3 sheep anti-mouse (Molecular Probes), for 45 minutes at room temperature. Cell nuclei were stained with DAPI 300 nM. Images of one sample per group were acquired by an Inverted Ti-E fluorescence microscope (Nikon).

*2.9 Statistical Analysis*

Results were expressed as Mean ± SEM plotted on graph. Statistical analysis was made by two-way ANOVA analysis of variance using GraphPad Prism software (version 6.0), with statistical significance set at $p \leq 0.05$.

**3. RESULTS**

*3.1 Physicochemical analysis*

Table I recapitulates the chemical and phase composition of B-HA and S-HA materials, overall porosity and specific surface area. In spite the overall porosity of the two materials is similar, there is a large difference in the SSA, ascribed to the presence of diffuse micro and nano-porosity in B-HA, that could be maintained thanks to the application of a process, based on heterogeneous chemical reactions, preventing the use of high temperature sintering and the consequent grain growth[26]. Such a method also allowed to avoid the crystal growth induced by high temperature treatments, thus retaining nanostructure and also $Mg^{2+}$ and $Sr^{2+}$ ions in the structure of B-HA, that could not be obtained, conversely, in the porous sintered S-HA.

**Table I.** Physicochemical and morphological data on B-HA and S-HA materials

|  | Phase composition | Ca/P (mol) | Doping ions (mol%) | | Overall porosity (vol%) | SSA (m²/g) |
|---|---|---|---|---|---|---|
|  |  |  | Mg²⁺ | Sr²⁺ |  |  |
| **B-HA** | Hydroxyapatite + β-TCP (~20%) | 1.58 ± 0.02 | 3.08 | 0.98 | 59 ± 3 | 12.55 |
| **S-HA** | Hydroxyapatite + β-TCP (~25%) | 1.60 ± 0.05 | - | - | 64 ± 4 | 4.85 |

*3.2 Morphological analysis*

The SEM image reported in Figure 1 shows the excellent structural similarity between B-HA scaffold and the wood selected as a template. Noticeably, the channel-like architecture of B-HA reproduces with great detail the typical architecture of the primary and secondary wood xylem, in turn well mimicking the osteon structure typical of compact bone[22,][31]. Figure 2 shows the typical multi-scale structure of B-HA characterized by at least three levels of pore size: micro (mean diameter of the order of 1 μm), meso (mean diameter of the order of 50 μm) and macro (mean diameter of the order of 300 μm) porosity.

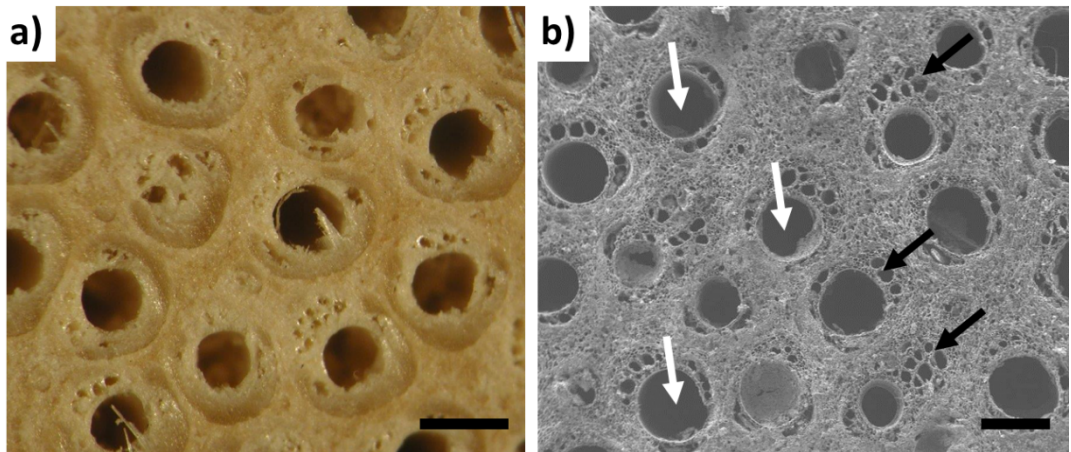

**Figure 1.** Comparison between a) the osteon structure typical of compact bone and b) the B-HA scaffold (top view). The metaxylem (white arrows) and protoxylem (black arrows) structures are evidenced. Scale bar: 200 μm.

Figure 2a shows the main channels of B-HA, typically 200-300 μm in diameter, replicating the metaxylem system that in the wood ensures the transport of the fluid from the roots to the upper part of the canes. Figure 2b highlights the smaller side channels, typically 40-50 μm in diameter, replicating the protoxylem tracheids, with its typical tubular elements of 5-10 μm organized as helical structures constituting the internal wall of the tracheid. Figure 2c shows the typical inner morphology of the macroscopic channels of B-HA, obtained by observing its transversal section by SEM. A cell structure is visible (on the left), well resembling the radial parenchyma (highlighted in Figure 2d), responsible of the lateral transport of fluids and nutrients to leaves. The right side of Figure 2c shows a section of the protoxylem channels departing from the scaffold top (see also Figure 2b). The insert in Figure 2d shows a high magnification SEM image of the building blocks constituting the B-HA scaffold, which are very thin lamellar nanoparticles, about 200-300 nm in size and 20-50 nm in thickness. Noticeably, the lamellae are closely intertwined, as they originated from dissolution-reprecipitation of a biomorphic $CaCO_3$ precursor during the final hydrothermal step of the biomorphic transformation process [26]. This allows B-HA to be a highly consolidated 3-D ceramic without any coalescence of the grains.

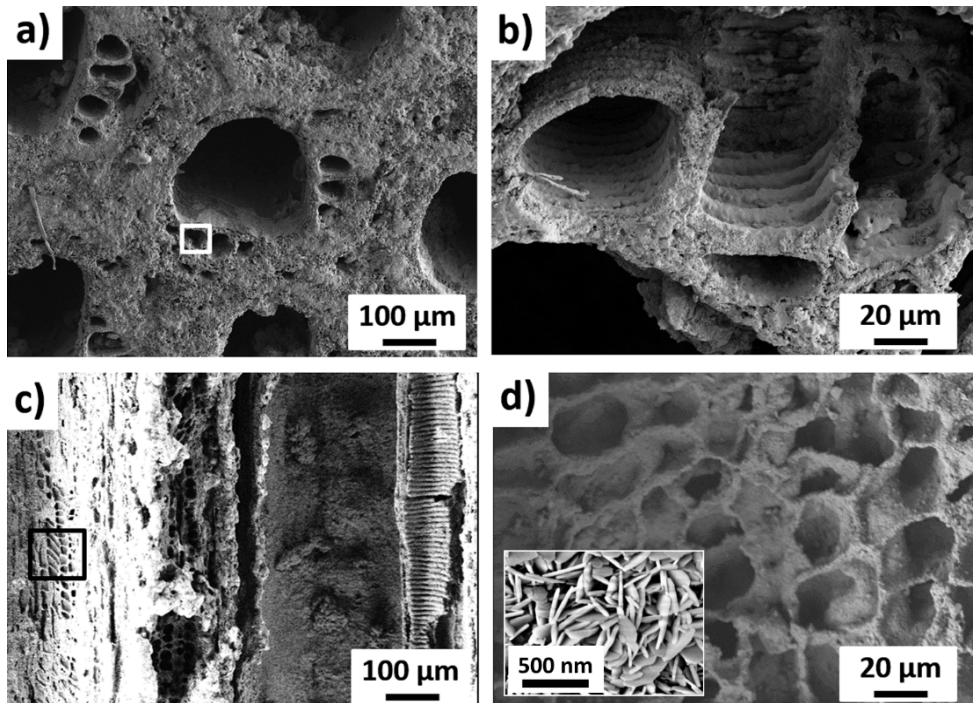

**Figure 2.** FE-SEM images at different magnifications showing the material microstructure at multiple scales. a) Top view of the main channels; b) detail of the protoxylem-like channels identified by the white box in a); c) lateral view of the main channel, highlighting its internal microstructure; d) detail of the parenchyma-like microstructure identified by the black box in c).

Figure 3 shows the microstructure of S-HA with rounded, interconnected macropores with size ranging between 200 and 500 µm. Contrary to B-HA, S-HA shows randomly organized porosity but, however, its pore interconnection can ensure very good wettability of the whole scaffold, similarly as occurs with B-HA scaffold.

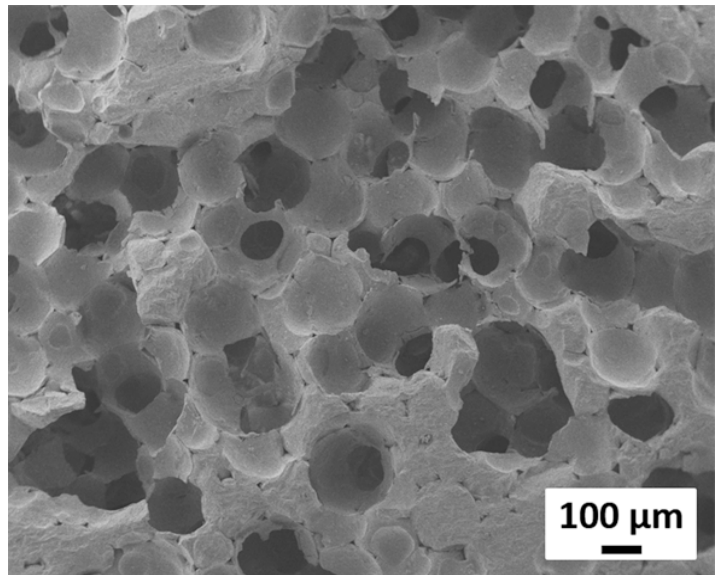

**Figure 3.** Morphology and pore architecture of S-HA scaffold.

*3.3 Mechanical characterization*

Elastic stiffness and strength of B-HA, estimated through uniaxial compression, three-point bending and ultrasonic tests, are reported in Table II. All values have been renormalized with respect to the mass density of 1.47 g/cm$^3$, corresponding to the average value of the density of the B-HA samples tested under compression. A typical stress-strain response obtained through a compression test is shown in Figure 4, for both B-HA and S-HA samples (both reported to the same mass density).

**Table II**. Results of the mechanical tests performed on B-HA and S-HA samples, where $\sigma_c$ denotes strength, $E_c$ the Young Modulus in compression and $\rho/\rho_0$ the ratio between the average density of the samples and the reference value of 1.47 g/cm$^3$ (average density of the B-HA samples tested under compression).

| Material | Compression | | | Ultrasonic (0.5÷1 MHz) | | Three-point bending | |
|---|---|---|---|---|---|---|---|
| | $\sigma_c$ [MPa] | $E_c$ [MPa] | $\rho/\rho_0$ [-] | $E_c$ [MPa] | $\rho/\rho_0$ [-] | $\sigma_f$ [MPa] | $\rho/\rho_0$ [-] |
| S-HA | 5.7±1.6 | 126±103 | 0.89 | 7307±404 | 0.36 | 5.00±1.38 | 1.01 |
| B-HA | 9.8±5.9 | 711±866 | 1 | 10814±1812 | 1 | 11.3±3.8 | 0.97 |

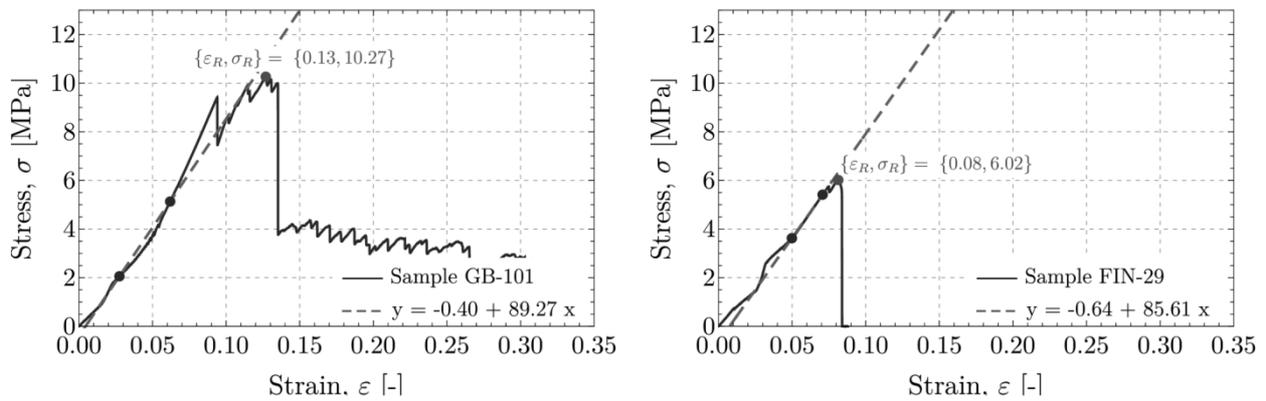

**Figure 4**. Stress-strain curves for the biomorphic apatite, B-HA (on the left), and sintered hydroxyapatite, S-HA (on the right), the latter renormalized to the reference density of 1.47 g/cm$^3$ of the former. The elastic stiffness represents the slope of the dashed straight line (drawn through the two blue dots) characterizing the mechanical behavior at small deformations. Note the superior strength and stiffness of B-HA.

The Ashby plots (or "charts") represent a useful tool for material selection in the design of a mechanical piece and provide a synthetic view of the most relevant material properties, when compared to other materials[32]. The Ashby charts mapping the Young modulus versus the strength of the material, and the Young modulus versus the porosity, are shown, respectively, in Figures 5 and 6 and compared with the conventionally sintered hydroxyapatite, S-HA and with other materials (whose data have been taken from Ashby's "Material and Process Selection Chart" and from [33, 34]. Samples of B-HA have been tested in the direction parallel (suffix '//') to the main fiber direction.

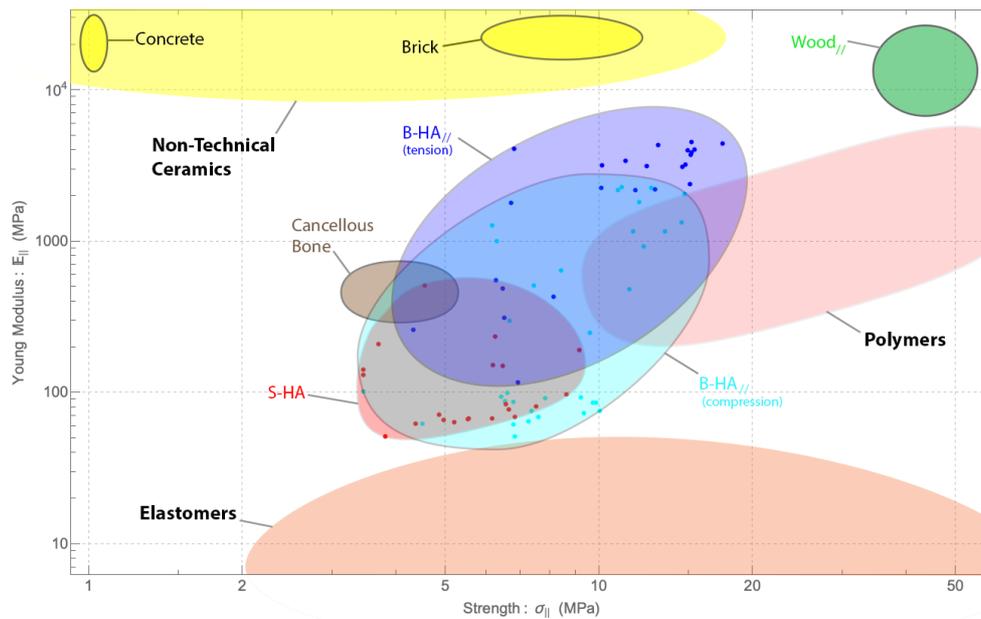

**Figure 5**. Ashby chart showing the mechanical behavior of B-HA, outstanding when compared to the sintered porous hydroxyapatite, S-HA (other materials reported for comparison were taken from Ashby's "Material and Process Selection Chart"). Note that B-HA shows a mean value of strength in tension superior than that in compression.

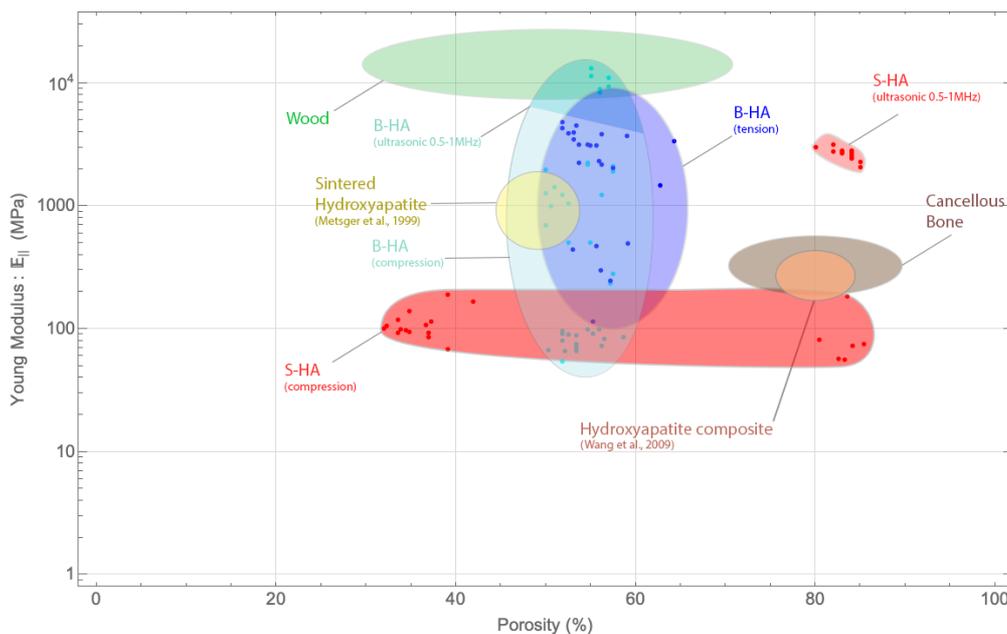

**Figure 6**. Ashby chart showing the mechanical behavior of B-HA, outstanding when compared to the sintered porous hydroxyapatite, S-HA (other materials reported for comparison were taken from Ashby's "Material and Process Selection Chart", when references are not reported).

The regions of the Ashby charts in Figs. 5 and 6 occupied by B-HA are typical of foams and polymers, but are usually virgin zones to ceramics, which is another proof of the unique mechanical characteristics of the B-HA. The hierarchical structure of B-HA inherits the nano- and micro-scale hierarchical optimization typical of its highly hierarchical parent material, the wood; this allows B-HA to express mechanical characteristics closer to that of the wood itself. The B-HA hierarchical structure is responsible for the excellent behavior in tension (where strength is higher than in compression) shown in Fig. 4 and Tab. II, but also for the difference between the elastic modulus exhibited at a small scale and that measured at a large scale, because in the former case the material response tends to become less affected by the structure and more by the inherent features of its basic component, namely the hydroxyapatite phase. Results on compression tests inside the SEM allow to investigate the scaffold toughness in terms of nucleation, growth, and coalescence of cracks, as shown in Figures 7-9.

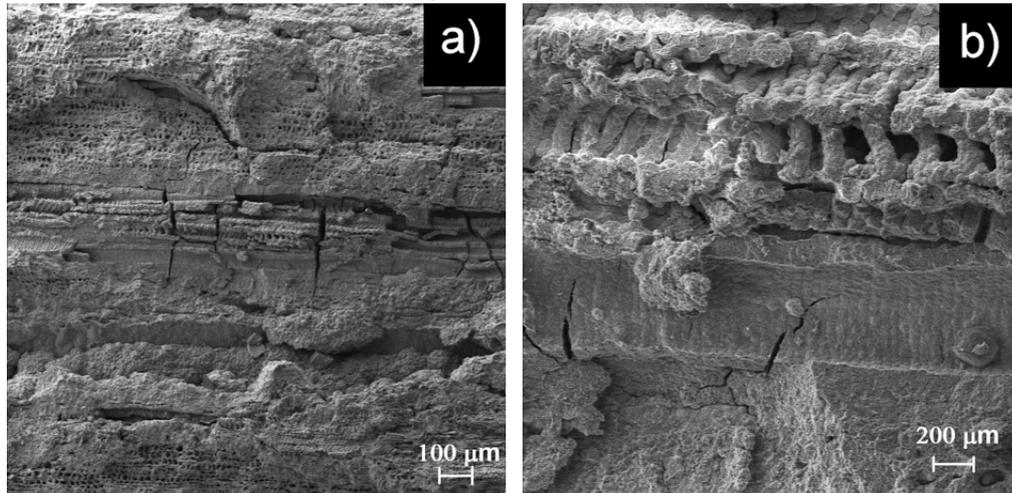

**Figure 7**. Post-mortem FE-SEM images at two different magnifications (150x magnification on the left, 1000x magnification on the right) of a sample of B-HA, tested under uniaxial compression parallel to the channels direction (horizontal direction in the images). The images highlight the tortuous crack pattern involving longitudinal and transverse cracks originated by the multi-scale porosity.

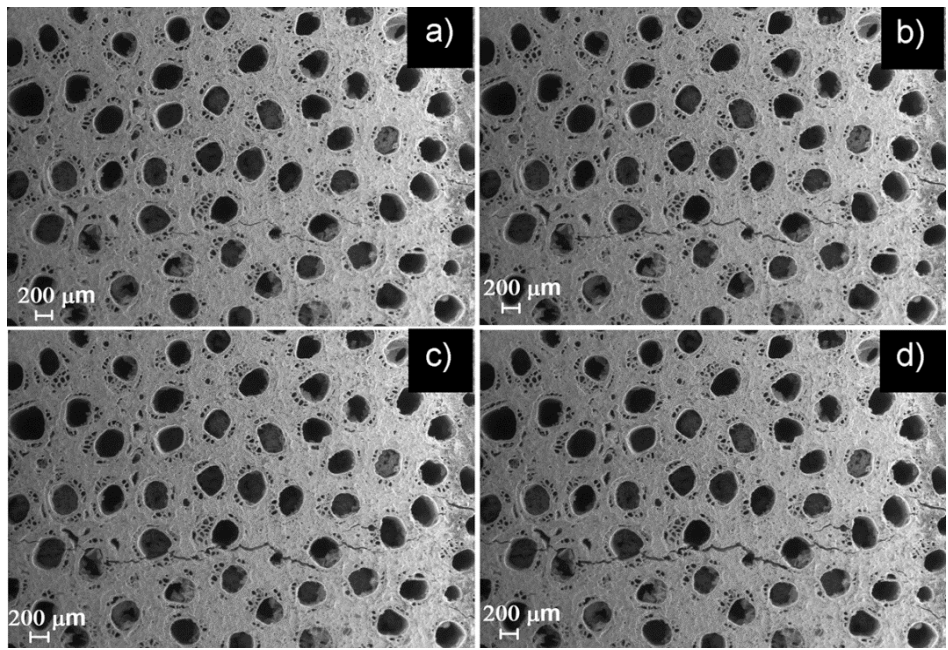

**Figure 8**. SEM images of a sample of B-HA, tested under compression in the direction orthogonal to the main longitudinal channels, for four different far-field imposed displacements whose magnitude is increasing from (a) to (d). A splitting crack nucleated from the channels and propagated in a tortuous manner due to the presence of micro-pores. The tortuosity is linked to the remarkably high ductility of the material.

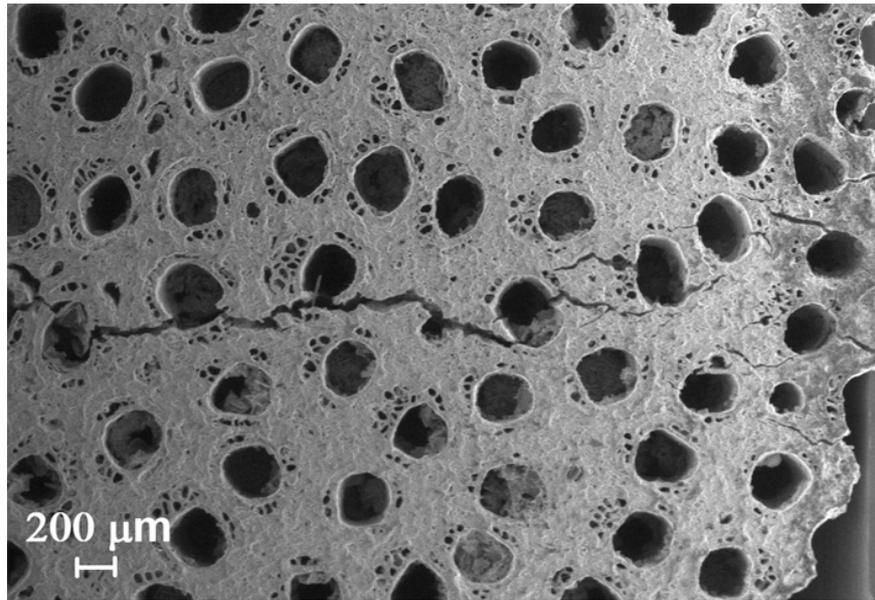

**Figure 9**. Post-mortem SEM image of a sample of B-HA, tested under compression in the direction orthogonal to the main longitudinal channels, showing the enhanced ability of the multi-scale porous microstructure to dissipate energy through multiple crack branching caused by the hierarchic pore distribution.

Fracture development is chiefly influenced by the loading conditions so that cracks are formed as a consequence of stress concentrations due to microbuckling-induced bending in Fig. 7 to a splitting mechanism generated by tensile stresses occurring at the pore surface. In both cases, an evident process of nucleation and growth of cracks is visible, which is hardly found in S-HA, characterized by abrupt failure. This fracture process is the signature of the superior ductility of B-HA, which is also related to the softening behavior shown in Fig. 4 on the left, absent in the figure on the right.

*3.4 Biological characterization*

As a confirmation of the instructive functions of B-HA at the end of the experiment, most of the aligned pores of the sample were found to be filled with cells and extracellular matrix secreted by the seeded cells (Figure 10). Moreover, the immunofluorescence analysis highlighted that hADSCs were secreting osteonectin, a glycoprotein that binds calcium, usually secreted by osteoblasts during bone formation initiating mineralization and promoting mineral crystal formation (Figure 11). hADSCs resulted positive to the anti-IBSP antibody confirming that cells were differentiating in osteoblasts. At the same time, HUVECs expressed VWF and PECAM1 as revealed by the immunofluorescence staining (Figure 11). These results confirmed the presence of both cells populations. From this qualitative analysis differences do not emerge between B-HA and the control S-HA. A deep quantitative investigation of the effectiveness of this dynamic co-culture system of hADSCs and HUVEC was performed by qPCR. The inductive effect of the B-HA scaffold was confirmed by the evaluation of the relative quantfication of the expression of RUNX2, IBSP, Coll1 and ALP (Figure 12) compared to the control. B-HA significantly up-regulates functional genes expressed in a mature osteoblast, Coll 1 and ALP ($p \leq 0.05$ and $p \leq 0.001$, respectively). Most fascinating, the level of VWF and PECAM1, the constitutive and functional genes of HUVEC cells, were found to be highly upregulated in the cells seeded on B-HA compared to CT.

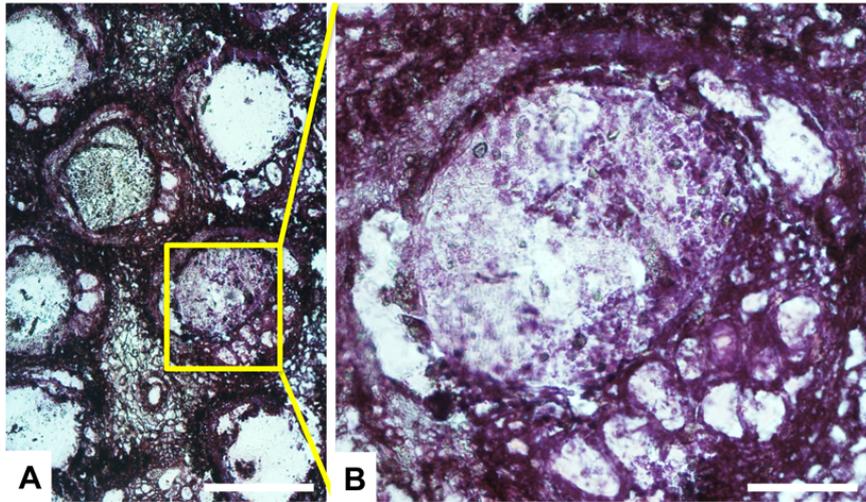

**Figure 10.** Hematoxylin and Eosin staining of B-HA sample. Scale bars: A 500 µm; B 250 µm.

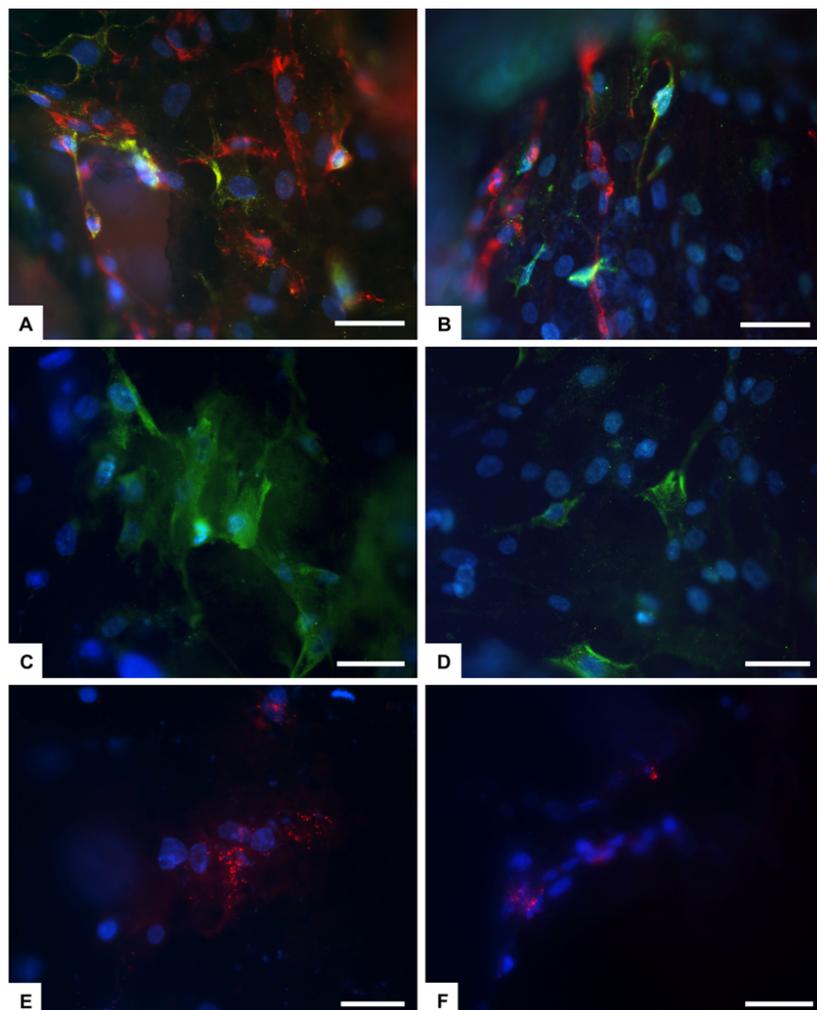

**Figure 11.** Immunofluorescent analysis: Osteonectin in green and PECAM1 in red (A, B); IBSP in green (C, D); VWF in red (E, F) and cell nuclei (in blue). A, C and E: B-HA group. B, D and F: S-HA, control group. Scale bars: A–F: 50 µm.

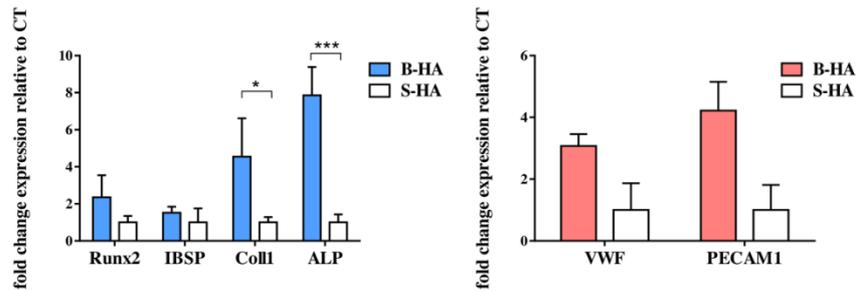

**Figure 12**. Relative quantification ($2^{-\Delta\Delta Ct}$) of RUNX2, IBSP, Coll1 and ALP expression as markers of osteogenesis differentiation for hADSCs, and VWF and PECAM1 as markers of the presence of HUVECs after 10+10 days of co-culture in bioreactor. The graphs report the mean and standard error of the samples with respect to the expression of control (*p≤ 0.05; *** p≤ 0.001).

## 4. DISCUSSION

The B-HA scaffold is obtained by an unconventional process based on heterogeneous chemical reactions carried out directly in the 3D state, described in a previous paper. This process uniquely allowed to obtain a 3D apatite scaffold retaining lattice substitution with $Mg^{2+}$ and $Sr^{2+}$ ions and endowed with a hierarchically organized nano-structure with multi-scale porosity from the macro to the nanosize[26]. As a control, a sintered ceramic material showing isotropic macro-scale porosity (S-HA) was chosen, made of hydroxyapatite and beta-TCP in similar amount as B-HA but without any doping ions because, following the recrystallization process induced by high temperature sintering, doping ions cannot be retained in the structure of hydroxyapatite phase [35].

B-HA shows a multi-scale hierarchical architecture built on strongly interacting nanosized lamellae forming complex tubular structures that, at higher size scales, constitute the walls of the wide and pervious channels typical of the original wood. In vegetable structures the flowing of the lymph is governed by a complex network of vessels, structured to facilitate the delivery of fluids and nutrients from the main channels to the thinner vessels and leaves. Particularly, wooden canes are characterized by a very efficient and safe fluid transport system, organized to prevent any embolism and ensure the delivery of nutrients up to the highest branches and leaves[31]. Such a structure was replicated with very good detail in B-HA: the original metaxylem and protoxylem structures were retained, including the annular or helical microtubular system sheathing the xylem walls (see Figures 2b and 2c). It can be devised that with this particular and unique microstructure, the B-HA scaffold can favor the fluid-dynamic mechanisms typical of natural vascular and lymphatic networks so as to promote biomimetic shear stresses, relevant also for the behavior of endothelial cell, active in the vascularization process [36-38]. Both the B-HA and S-HA scaffolds can support continuous fluid flow; however, whereas S-HA exhibits an isotropic macroporosity with tortuous pore organization (see Figure 3), B-HA combines highly oriented channels with a hierarchically organized interconnected nano- and micro-porosity that may favour the fluid to flow more similarly to blood circulatory systems.

The hierarchical architecture of B-HA strongly affects its mechanical properties, in terms of:
i) a mean value of tensile strength greater than mean value of compressive strength (Table II, Figs. 4 and 5) and both superior to values corresponding to S-HA. The fact that a material displays a tensile strength higher than its compression strength is typical of fibrous materials and in particular of wood, so that this property is believed to be inherited from the ligneous matter from which the ceramics has been obtained;

ii) a mean value of Young's modulus, measured parallel to the main channels, greater than that of S-HA (Table II, Figs. 5 and 6), a property which makes the mechanical behaviour of the material closer to the elastic properties of bone. From the mechanical point of view, B-HA is thus an optimal candidate for bone repairing as its Young modulus is found similar to the modulus of bone obtained with the mixture rule from the values corresponding to compact and cancellous bone. The rule of mixture is used to predict properties of composite materials, and it consists in the weighted average of a mechanical property over the volume fractions of its constituents:

$$E_{equivalent} = E_{cancellous} \frac{V_{cancellous}}{V_{total}} + E_{compact} \frac{V_{compact}}{V_{total}}$$

The above rule allows to estimate the overall behaviour of a composite material. For flat bones, such as human's cranial bone for example, using the aforementioned equation leads to: $E_{equivalent} \cong 2000$ MPa, which is a value compatible with the results obtained for B-HA.

iii) a ductility unexpectedly high for a ceramic material, so that Figs. 4 evidences an important softening behaviour, while Figs. 7-9 evidence a surprising fracture tortuosity for ceramics;

iv) the Young modulus vs. strength Ashby plot in Fig. 5 reveals that B-HA occupies a zone usually not covered by ceramics, another proof of the special characteristics of this material.

The outstanding mechanical performance of B-HA, unusual for calcium phosphate-based ceramics, can be ascribed to the maintenance of strongly intertwined nanosize building blocks, in association with the multi-scale pore hierarchy that allows progressive fracturing of the scaffold and delayed failure, in comparison with the brittle character of sintered ceramic bodies. Such a behaviour, uncommon among ceramic materials, is made possible by the use of a very unconventional method for B-HA fabrication, based on heterogeneous chemical reactions occurring between a solid biomorphic template and a reacting gas, followed by a hydrothermal process forming the final B-HA scaffold [26]. On one hand, the process allowed the formation of highly consolidated 3D ceramic formed by nanotwinned lamellar nanoparticles rather than by conventional grain coalescence, therefore preventing failure phenomena related to the presence of grain boundaries. On the other hand, thanks to the fine control of the reaction kinetics, phase transformation into the final hydroxyapatite phase could occur directly in the 3-D state without altering the fine microstructural details of the original wood template.

The excellent bending strength and stiffness of B-HA is a feature relevant for applications in bone regeneration. Given the unique 3-D structure, mimicking the complex architecture of compact bone, and the mechanical performance, including excellent strength and ductility for hydroxyapatite-based ceramics, the B-HA scaffold is promising for enhanced ability to activate bone cell mechanotransduction which is a cellular process crucial in bone regeneration, and particularly relevant in load-bearing regions[39].

To better recapitulate the heterogeneity of the new bone formation process and obtain more significant results, in this study hADSCs were seeded on the U-Cup bioreactor, able to establish a long term 3D co-culture systems [40, 41] for 10 days in order to create a biomimetic microenviroment that could promote the HUVECs adhesion on the scaffold surface. Then HUVECs were added to the culture, and the obtained co-culture were analyzed after 10 additional days. Therefore, the proposed co-culture setup summarizes essential features also of the nature of cellular communication so that an useful *in vitro* model of vascularization which remains one of the major restrictions in regenerating large defects[42], is provided. Taking into account that significant differences in endothelial cell behavior were previously attributed to the matrix dimensionality[43], the presented results, although preliminary, are significant as the study was conducted in a biomimetic 3-D environment, and are indicative of a general improvement towards new bone and vascular development given by the biomimetic features of the B-HA scaffold, enhanced with respect to sintered porous apatites. In our previous study we showed the ability of a B-HA scaffold to enhance the viability of human mesenchymal stem cells than could easily colonize the scaffold itself, and to induce an upregulation of several genes involved in osteogenesis[26]. The exchanging $Mg^{2+}$ and $Sr^{2+}$ ions in cell culture media played a fundamental role in the enhancemnt of the osteogenic activity[26]. Here, a perfusion co-culture sistem were established to demonstrate that the B-HA scaffold enhances the activity of mesenchymal and endothelial cells, attested by the overexpression of relevant genes such as Coll1, ALP, VWF and PECAM-1. This suggests that a microenvironment with appropriate physico-chemical and mechanical features given by inherent scaffold properties can promote cell-cell interaction that is fundamental in tissue regeneration process. Several reports, in fact, suggested that *in vitro* both endothelial and mesenchymal stem cells behavior are synergistically and positively affected by mutual presence[26, 44-47].

In this study the PECAM1 and VWF genes upregulation in HUVEC cells induced by B-HA suggest that the presence of $Mg^{2+}$ and $Sr^{2+}$ ions in the culture sistem can be beneficial also for the endothelial cells activity as already demonstrated [48, 49],even though deeper investigation is required to unveil the fine mechanisms of the biological cascade induced by such ions. Another characteristic that we consider a key factor in influencing the cell behavior is the porous channel-like structure of the scaffold [36-38]. It is well known that the endothelial cells can be affected by fluid shear stresses[50] which in turn are governed by the pore geometry and interconnection of the seeded 3-D substrate.

## 5. CONCLUSIONS

A 3-D ceramic material (B-HA) previously obtained by an unconventional fabrication approach based on biomorphic transformation of a natural wood, a process not requiring the use of high temperature sintering for the final ceramic consolidation, revealed its outstanding architecture resulting in large channels interconnected with smaller tubules that reproduce in detail the lymphatic system of the original wood template. From the mechanical viewpoint, B-HA showed superior mechanical properties when compared to sintered apatite scaffolds with similar, but isotropic, porosity extent, in particular higher stiffness and strength, both displaying directional properties which mimic the directional properties of wood and bone. Moreover, the average strength in tension of B-HA was found higher than that in compression, a unique feature resulting from the hierarchical arrangement of channels and pores across the size scales which can lead to microbuckling-induced bending up to splitting, generated by tensile stresses occurring at the pore surface. Remarkably, B-HA displayed a larger deformation level at failure when compared to S-HA, as a result of tortuous crack paths allowing for a much greater energy dissipation during fracture and resulting in high ductility, unexpected for a pure ceramic material. Co-culture tests conducted in bioreactor revealed significant upregulation of various genes involved in osteogenesis and angiogenesis process. Hence, the 3D biomimetic architecture of B-HA and damage-tolerant mechanical performance make B-HA promising as a scaffold able to enhance bone and vascular regeneration, particularly in long bone segments, where the lack of vascularization with the current implants is a major concern.

## ACKNOWLEDGEMENTS

This work was supported by the Italian Ministry of Education, University and Research, under the project PRIN 2015 'Multi-scale mechanical models for the design and optimization of micro-structured smart materials and metamaterials' 2015LYYXA8-006. This work was also supported by the National Group of Mathematical Physics (GNFM-INdAM).